\documentclass[useAMS,usenatbib]{mn2e}
\usepackage{graphicx}

\usepackage{amsmath}

\newcommand{\aap}{A\&A}
\newcommand{\mnras}{MNRAS}
\newcommand{\apj}{ApJ}
\newcommand{\aj}{AJ}

\newcommand{\apjs}{ApJS}
\newcommand{\araa}{ARA\&A}
\newcommand{\nat}{Nature}
\newcommand{\sci}{Science}

\catcode`\@=11
\def\gsim{\ifmmode{\mathrel{\mathpalette\@versim>}}
    \else{$\mathrel{\mathpalette\@versim>}$}\fi}
\def\lsim{\ifmmode{\mathrel{\mathpalette\@versim<}}
    \else{$\mathrel{\mathpalette\@versim<}$}\fi}
\def\@versim#1#2{\lower 2.9truept \vbox{\baselineskip 0pt \lineskip
    0.5truept \ialign{$\m@th#1\hfil##\hfil$\crcr#2\crcr\sim\crcr}}}
\catcode`\@=12
\def\msun{\hbox{$M_\odot$}}

\def\yr-1{\hbox{${\rm yr}^{-1}$}}

\def\t9{\hbox{$t_9$}}

\def\m*{\hbox{$M_{\rm stars}$}}

\def\ho{\hbox{$H_\circ$}}
\def\h50{\hbox{$\ho /50$}}

\begin{document}

\title{On the Angular Momentum History of Galactic Disks} 
\author[Alvio Renzini]{Alvio Renzini$^{1}$\thanks{E-mail: 
alvio.renzini@inaf.it}\\ 
 $^{1}$INAF - Osservatorio
Astronomico di Padova, Vicolo dell'Osservatorio 5, I-35122 Padova,
Italy}

\date{Accepted March 24, 2020; Received February 7, 2020, in original form}
 \pagerange{\pageref{firstpage}--\pageref{lastpage}} \pubyear{2002}

\maketitle
                                                            
\label{firstpage}

\begin{abstract}
The stellar mass, size and rotational velocity of galactic disks all grow from redshift $\sim 2$ to the present by amounts that are estimated from observationally derived scaling relations. The product of these three quantities, the angular momentum of stellar disks, is then estimated to grow by a remarkably large factor, between $\sim 20$ and $\sim 50$, whereas other evidences suggest a more moderate increase. This requires that the specific angular momentum of the accreted gas should systematically increase with time while remaining co-rotational with the disk over most of the last $\sim 10$ Gyr. Thus, the baryonic gas vorticity of the circumgalactic medium appears to emerge as a major driver in galaxy evolution, and this paper is meant to attract attention on the sheer size of the angular momentum increase and on the need to explore to which extent this can be observed in nature and/or in simulations.

\end{abstract}


\maketitle

\section{Introduction}
\label{intro}
Most galaxies in the local Universe  are rotationally supported and since they grew from small seeds to giant dimensions their stellar angular momentum (hereafter AM, or $J_*$) must have grown accordingly. Yet, the AM evolution of individual galaxies remains largely unexplored.
In a recent paper (\citealt{peng20}, hereafter PR20) we have argued that the stellar AM of galaxies that remain star-forming all the way to the present should increase by a very large factor over the last $\sim 10$ Gyr, i.e., since the Universe high-noon at redshift $\sim 2$ (i.e., when the majority of galaxies settle into ordered rotation). The traditional approach to the origin of the AM of galaxies has been essentially theoretical, top-down from first physical principles. This is not the approach of this paper. Here instead the reasoning proceeds bottom-up, from the empirical scaling relations for star-forming galaxies to their  implied history of AM growth, to finally asking what physical conditions may have ensured such growth, including the extent to which they may have been established in relevant simulations.

The size of the AM increase follows directly from the scaling relations for galaxies that remain close to the main sequence (MS) in the course of their evolution, as the AM  of a galaxy is:
\begin{equation}
J_*\propto M_*R_{\rm h}v_{\rm rot},
\end{equation}
where the three factors represent the stellar mass, the half-mass radius and the rotational velocity, respectively. 

We start by briefly recapping the PR20 estimate of the AM evolution, though with slightly different numbers.
From the specific star formation rate (sSFR) of MS galaxies evolving as $\sim (1+z)^{2.8}$, PR20 argue that on average star-forming galaxies increase their stellar mass by a factor $\sim 10$ since $z\simeq2$, thought the
size of such increase is very sensitive to the actual zero point of the MS relation \citep{renzini09}. Moreover, the extent to which the MS {\it bends} at high stellar masses still differs substantially from one study to another \citep{ppop19}, which adds further uncertainty to the result of the MS integration.     
An alternative way of estimating the typical mass increase of MS galaxies rests on the consideration that the cosmic stellar mass density (in $\msun{\rm cMpc}^{-3}$) increases by a factor $\sim 4$ since $z\simeq 2$ \citep{madau14}, whereas the shape of the mass function of star-forming galaxies does not change much over the corresponding time interval (e.g., \citealt{peng10,ilbert13,muzzin13}). Thus, an increase by a factor of $\sim 4$ represents a strict lower limit to the mass increase of individual MS galaxies. Indeed, at $z\sim 2$ most galaxies are still star forming near the MS whereas at $z=0$ over 50 per cent of stellar mass resides in quenched galaxies. Since only star-forming (MS) galaxies contribute to the increase of the cosmic stellar mass density, each of them has to increase its stellar mass by more than a factor of 4, so to compensate for those galaxies that cease to contribute as they quench. Thus, a fair estimate of the average stellar mass increase is by a factor $\sim 8$, or just a little less than that, with a decreasing trend for the most massive galaxies.

As far as galaxy sizes are concerned, PR20 adopt the scaling
\begin{equation}
R_{\rm h}\propto (1+z)^{-1}M_*^{0.2}
\end{equation}
 \citep{newman12,vdw14,mosleh17} for the half-mass radius of disks, hence a size increase by a factor of $\sim 3$ since $z=2$, at fixed stellar mass, plus the effect of the mass increase by a factor of $\sim 8$ implying  an additional increase by a factor $\sim 1.5$, making a total of  $\sim 4.5$ times for the typical size increase of individual disks. Concerning the third factor in $J_*$, the rotational velocity $v_{\rm rot}$ of individual MS galaxies increases by a factor $\sim 1.5$ since $z\simeq 2$ \citep{simons17}, an effect that was mentioned, but neglected, in PR20. In summary, the resulting AM increase is by a factor $\sim 8\times 4.5\times 1.5\simeq 50$. The dominant factor is the mass increase, which is perhaps the less constrained one by direct observations. If one adopts the factor of 4 as a lower limit, 
then the increase in AM is still by a factor $\sim 20$. Clearly, further refining the scaling relations for the three quantities in Equation (1) is of great importance for a reduction of  the uncertainty in the size of the AM evolution. Moreover, it is worth emphasizing that not only the total AM of disks increases with time, but so does also their specific AM (sAM), as from Equation (1) $j_*=J_*/M_*\propto R_{\rm h}v_{\rm rot}$ and this latter product increases by a factor $\sim 6$ since redshift 2, if the adopted scaling is correct. This factor applies to the sAM of the whole stellar disk, therefore implying that the sAM of the gas being accreted (and turned into stars) must secularly increase by an even much larger factor over the same time interval.  

As emphasized in PR20, this macroscopic increase in the AM of galactic disks requires a systematic increase with time of the AM of the gas being accreted from the environment, coupled with the accretion itself being nearly co-planar and co-rotating with the disks themselves. Thus, the AM stored in the circumgalactic medium (CGM),  actually 
of the fraction
 of it which is actually accreted, drives the growth of the disk. This rises a series of questions: is all this really taking place in nature? What empirical evidence d
 oes exist for it? What is the size and AM distribution of the CGM domain having fed local galaxies? Do simulations offer any hint in this respect? This paper does not offer answers, it is rather meant to attract the attention on the sheer size of the AM growth, hence on its implications for the mechanisms that promote the growth of disks. 

Yet, this large increase of the sAM is not expected in the frame of canonical dark matter theory, that predicts the sAM of disks to evolve as $\sim (H(z)/H_\circ)^{-1/3}$, corresponding to a just $\sim 40$ per cent increase since $z=2$, in agreement with the results of the integral field spectroscopic survey of $z\sim 1-3$ galaxies discussed by \cite{burkert16}. Moreover, at fixed stellar mass, no sAM evolution  since $z=1$ is found by \cite{maresco19} in a subsample of galaxies from same survey, though these galaxies may have been originally selected for being among the most {\it mature} disks at this redshift (N. M. F\"orster Schreiber, private communication).  On the other hand, the radial growth rate of local disks is estimated to be $\sim 0.35$ times that of the stellar mass \citep{pezzulli15}, that if constant over time would imply a size increase by a factor $2-3$ for our estimated mass increase since $z=2$. Moreover, even the empirical size scaling with redshift as from Eq. (2) may not imply the full corresponding growth of individual galaxies, if smaller, denser disks were more prone to quench than the bigger ones \citep{vkk15}, hence implying an increase in the average size of the surviving, star-forming disks\footnote{But see \citep{lilly16} for a different interpretation of the apparent correlation between quenching and galaxy density}.
Thus, by no means univocal evidence has yet emerged concerning the evolution of the sAM.

\section{Do Galaxy Filaments drive the growth of disks?}
\label{fili}
This large increase of the AM of disks could not happen were the gas accretion chaotic. It requires instead a long term coherence to maintain near co-planarity and co-rotation while the AM of accreting gas has to increase with time. Thus, what we need is a persistent structure around galaxies with a naturally built-in organization that must be automatically conducive to the required co-planarity and co-rotation of the gas inflow with secularly increasing AM. Galaxy filaments appear to be obvious candidates for making all this happening, as filaments naturally arise from the gravitational instability of the (dark) matter distribution, hence setting preferential directions.

However, for filaments to do the job their structure should satisfy certain conditions that may or may not be established in nature. Qualitatively, we may expect the baryonic gas to rotate around the axis of a filament, as it is attracted by the filament gravitational pull,  while roughly conserving the AM it may have acquired from tidal interactions with 
other forming structures nearby. A gradient in specific AM, perpendicular to the filament, will also naturally arise, as the more distant material is expected to have experienced stronger tidal interactions with its surroundings, hough the cross section of filaments is typically of the order of a Mpc \citep{tempel14}, much larger than galaxies. Such a rotating baryonic cylinder would quite naturally administer to galaxies the raw material for their growth, with increasing AM, hence growing disks whose rotational vector should align with the filament. This would  indeed be the predicted signature of such a scenario, that can be subject to test in observations and simulations.

Since \cite{peebles69}, tidal interactions are seen as the origin of the AM of galaxies. This tidal-torque theory has been widely explored (including its limitations) in particular with N-body simulations, to infer the spin (AM) of dark matter halos (e.g., \citealt{porciani02,hahn07}) and their tendency to form bigger and bigger spheroidal halos. Encouragingly, \cite{laigle15} and \cite{codis15} find that the resulting  vorticity of the dark matter tends to align with the filaments.  Yet, what matters here is the vorticity and AM of the baryons and their tendency to dissipate and form (thin) disks. 

However, the observation of the dynamical configuration of the gas in filaments is largely beyond our current capabilities. Yet, in principle this  is thoroughly {\it observable} in simulations. The baryon inventory in cosmic knots, filaments, sheets and voids in one such simulation has been recently illustrated by \cite{martizzi19}. They provide the distribution of the various gas phases among the mentioned structures also at different redshifts, but do not extract from the simulation how the gas moves within the filaments. This may come in a later paper by the same team.

Galaxy alignment (or lack of) with respect to filaments has been instead quite widely explored, both from direct observations and in simulations.  In simulations, \cite{dubois14}  find indeed that at $z=1.8$ the spin vector of galaxies tend to align with filaments, and even more with the vorticity vector of the baryonic gas, as expected in this picture. But the signal is very week, with only a $\sim\!\! 10$ per cent excess of alignments  with respect to random, with the signal decreasing with cosmic time down to $z\sim 1.2$ and vanishing  altogether by $z=0$ in the same simulation \citep{codis18}.  Moreover, the spin tends to orient orthogonal to the filaments above a critical mass ($\sim 3\times 10^{10}\,\msun$), as a result of merging. In another simulation \citep{veena19} preference for perpendicular alignment at all masses is found, though the simulated cosmic volume is considered
insufficient. In essence, it appears that simulations produce some galaxy spin-filament alignment, but too weak to claim support for our ansatz of gas filament global vorticity being  the prime driver for the growth of galactic disks.

On the observational side, on SDSS, hence $z\simeq 0$ data, \cite{tempel13a} and \cite{tempel13b} find the spin vector of spirals to be preferentially parallel to filaments whereas that of ellipticals is preferentially orthogonal to them. The effect is small, though of the nearly $\sim\!\! 10$ per cent size as that \cite{dubois14} will find at $z=1.8$ in their simulations, but \cite{codis18} fail to find at $z=0$. More recently, and still using SDSS data, \cite{chen19} find the major axis of galaxies to be preferentially parallel to filaments, which implies a spin vector perpendicular to them, apparently at variance with the Tempel et al. findings. \cite{krolewski19} find no clear alignment signal with the galaxy spin being derived from MaNGA integral field spectroscopy, whereas \cite{blue19} find a prevalence of spin-filament alignments in a sample of ten late type galaxies whose spin direction is derived both for the stellar and the gas kinematics. These trends, modest spin-filament alignment for spiral/low mass galaxies and modest orthogonal alignment for elliptical/high mass galaxies is also found in the SAMI galaxy survey \citep{welker20}, basically in agreement with (most) previous studies over real data and simulations.

Thus, the bottom line is that, at all redshifts below $\sim\!\! 2$, $\sim\!\! 90$ per cent of disk galaxies are randomly oriented  with respect to their closest filament. It follows that the global vorticity on the scale of  filaments does not seems to play a major role on the ordered growth of disks and their AM. 
So, if not  the filaments, what else?
Well, before abandoning the idea, a closer look to some of the above studies is in order. For example, \cite{dubois14} postulate that {\it filaments have no polarity} and do not distinguish between spin vector orientations in one or an opposite quadrant with respect to filaments\footnote{Note that \cite{laigle15} find that segments of the (dark matter) filaments do exhibit polarity.}. In other words, the measured angle $\theta$ between the galaxy spin and the filaments is let to vary only between 0 and $\pi/2$, not between 0 and $\pi$, and cos$\,\theta$ between 0 and 1, rather than between $-1$ and 1.  So, in principle all the simulated galaxies could spin  with their vectors pointing in the same half space with an average $\theta$ just a little smaller than 45$^\circ$, which would correspond to a much stronger coalignment. 
Indeed, the same authors relax the no-polarity assumption, allow $\theta$ to vary between 0 and $\pi$, and find a stronger co-alignment between the galaxy spin and the vorticity of the gas, with parallel spins being $\sim 50$ per cent more frequent than antiparallel ones. This figure refers to the simulation at $z=1.8$, that observationally corresponds to still an incipient phase in the establishment of orderly, rotationally-supported disks. Thus,  it would be  interesting to explore  in the simulation to which extent the degree of this co-alignment increases towards lower redshifts, because observationally it is at lower redshifts that orderly-rotating disks become the dominant MS population and evidence exists for co-rotation of the CGM \citep{crystal19b,crystal19a,zabl19}. Thus, a direct role of filaments in organizing baryon vorticity cannot be excluded at this stage, but the suspicion is that intra-filament  galaxy-galaxy tidal effects may mess up the picture.

\section{Discussion}

It is widely recognized that AM and its accretion must play an important role in galaxy evolution (e.g., \citealt{danovich15,stewart17}),  but the sheer size of the AM increase,
possibly as large as a factor of $\sim 20\!-\!50$ since $z=2$, may have not been fully appreciated. The accretion of gas into galaxies  is generally seen as indispensable to sustain their star formation, given the short gas depletion times (e.g., \citealt{tacconi18} and references therein) and from the early days of theoretical galaxy formation attention has more often  focused on the thermodynamical aspect of accretion, i.e., on the required cooling of the CGM (e.g., \citealt{dekel09}). On the other hand, simulations paying special attention to the role of AM may have covered only a relatively narrow  interval of cosmic time, hence recovering only a fraction of the total AM growth (e.g., \citealt{danovich15}).

In the simulation discussed by \cite{pillepich19} disk radii grow by a large factor since $z\sim 2$, especially for $M_*>10^{10}\;\msun$, (see their Figure B1), that, coupled with the large expected increase in stellar mass, implies indeed a large increase of the AM of the stellar disks. Again, this means that, to some extent,  the simulation does produce a CGM with sufficient vorticity to drive the disk growth together with its AM,  in a roughly consistent fashion with what indicated by the scaling relations, as quantified in Section~1. However, these authors lament that ``no quantitative analysis of the spatially averaged or map-based internal kinematics of star-forming galaxies within large uniform-volume simulations exists". Not to mention the same kind of analysis for the CGM over a volume at least as large as that from which all the baryons having fed a galaxy came from i.e., a fraction of  the virial radius of the host halo, given that at most $\sim 1/3$ of the baryons in a halo are converted into stars \citep{behroozi10}. The simulated data exist, but they have not been {\it observed} yet.  Still, this simulation produces rotational velocities that decrease with time, admittedly at tension with the observational result in which they appear to increase with time \citep{simons17}\footnote{This may result from the simulation overestimating the size growth.}. 

More recently, the simulations of disk galaxy evolution by \cite{buck20} quite effectively illustrate the appearance of a spontaneous symmetry breaking between co-rotating and counter-rotating stellar orbits starting to take place at $z\sim 2$ (their Figure 10). From redshift $\sim 4$ down to $\sim 2$ stellar orbits are characterized by high velocity
dispersion and strong vertical motions, with nearly as many stars rotating in one direction as in the opposite direction, which may correspond to the formation of a bulge. At lower redshifts, especially at $z\sim  1$, the disk then grows very rapidly along with its AM, without further appearance of counter-rotating stars. Thus, persistent co-rotation of the accreted gas must be realised in this simulation and it would be interesting to extract from it the corresponding history of the AM stratification of the involved circumgalactic gas and of the stellar disk. In other words, the initial chaotic assembly is then superseded by smooth accretion of co-rotating gas, with this beginning at $z\sim 2$ being in nice agreement with the observed emergence of rotationally supported disks around this epoch (e.g., \citealt{simons17,forster18,ubler19}).

According to an early postulate, baryons would share the same sAM of their host dark matter halo. This is not what found in several independent hydrodynamical 
$\Lambda$CDM simulations analyzed by \cite{stewart17b} for the redshift range $1<z<3$, where the sAM of the cold, effectively accreting CGM is found to systematically exceed by 4--5 times that of the host halo. Moreover, the simulations show that the cold accreting gas is typically co-planar and co-rotating with the central galaxy and slowly inspiraling towards it, i.e., exhibiting all the features that are required for the secular increase of the AM of galactic disks that is demanded by the galaxy scaling relations. Still, it remains to be established whether this remarkable, but still qualitative agreement with the expectations from the scaling relations will turn quantitative and extended to $z=0$. 

Thus, we still don't know what is the AM growing factor of the simulated galaxies and how it compares to that predicted from the scaling relations (the $\sim 20-50$ factor).
Indeed, at stake is a better understanding of the intimate workings of galaxy evolution, with AM --and the history of its acquisition-- being a prime mover for the growth and evolution of galaxies. The mere existence of the main sequence of star forming galaxies represented a change of paradigm, with emphasis shifting from merging
to quasi stationary gas inflow as the main driver. In this frame, the gas-regulator model offers a simple  mechanism to smooth fluctuations in sSFR driven by fluctuations in the gas accretion rate, thus offering an explanation for the existence of the main sequence (e.g., \citealt{lilly13}). However, this gas does not carry just fuel to sustain star formation: it carries also AM. If it were to carry too much AM, then a galaxy may even starve and quench, as suggested in PR20. If it does not carry enough AM (or it comes with the wrong sign), what can be produced is perhaps something like a {\it blue nugget} \citep{dekel14b}, that may soon quench as well. Possibly, only if the acquired AM  keeps within certain limits then a successful disk will be produced, with baryonic gas vorticity on a circumgalactic scale acting as a {\it natural selection} process. Still, the conditions leading to success must be relatively widespread in nature, given that most galaxies spend a major fraction, if not all their lifetime, close to the main sequence and following the corresponding scaling relations. Given the enduring difficulty to fully probe empirically the CGM and its kinematical history, some light on this problem may be shed by {\it observing} it in the existing simulations, checking to what extent baryon vorticity is what determines the growth or the starvation of galaxies. Since quite many years  the  star formation history (SFH) of galaxies has been at the focus of both observation and theory. What needs to come on focus now is their angular momentum history (AMH), perhaps the next challenge for galaxy evolution studies.

\section*{Acknowledgments}

I wish to thank  Emanuele Daddi, Natascha F\"orster Schreiber, Mauro Giavalisco, Simon Lilly, Yingjie Peng, Gabriele Pezzulli, and Giulia Rodighiero for useful conversations on these matters and acknowledge support from the INAF/PRIN-SKA 2017 "ESKAPE-HI" grant.

\label{lastpage}

\end{document}